\newcount\pageformat\pageformat=2  
\newcount\pssize

\ifnum\pageformat=2
 \documentclass[aps,prb,twocolumn,floatfix,showpacs,amsmath,amssymb]{revtex4}
\else
 \documentclass[aps,prb,preprint,floatfix,showpacs,amsmath,amssymb]{revtex4}
\fi
\usepackage{graphicx}

\begin{document}
\title{On Excited States of Deuteron Nucleus}

\author{B. F. Kostenko} \email{bkostenko@jinr.ru}

\affiliation{Joint Institute for Nuclear Research, Dubna\\
141980 Moscow region, Russia}

\author{J. Pribi\v{s}}

\affiliation{Technical University, Ko\v{s}ice,  Slovak Republic}

\date{\today }

\begin{abstract}
For a long time it was known that deuteron, as a weakly coupled
nucleon pair, has no excited states. However, A.M. Baldin et al,
commenting results of the first physical experiment with accelerated
nuclei at JINR synchrophasotron, assumed as far back as in 1979 that
one of peaks in a differential cross-section may arise due to an
"excited state of deuterium". We have established that one of the
peaks in the cross-section may  be  explained indeed in this way and
corresponds to the dibaryon reported by WASA-at-COSY Collaboration.
Another peak in the same region is  interpreted most likely by
interference  of several  $N^*$-resonances, and this possibility was
also mentioned in the paper by A.M. Baldin et al. Further
experimental studies based on modern experimental facilities and
more abundant statistics are necessary to verify these observations.
\end{abstract}


\maketitle

\section{\label{sec1}Introduction}
Recently a proposal of QCD investigation at high density and low
temperature, complementary to the  heavy nuclear collisions, was
suggested \cite{Kostenko1, Kostenko2}. The proposal is based on the
fact that a large number of nucleons in the interaction region is
not  necessary for the phase transition to occur, and only a change
of the vacuum state should be initiated by some experimental
environment. In particular,  observation of multi-baryons (MB) may
be a direct evidence of  phase transitions in small nucleon systems.
Separation of a MB mass from the secondary particle background is
feasible if the MB decay width is narrow enough.  That requires the
excitation energy of  MB produced should be low. For this purpose,
it is reasonable to select only those experimental events in which
the MB creation is accompanied with a high momentum particle, taking
away an essential part of the energy from the interaction region --
method of cumulative particle \cite{Kostenko2}. In this paper, we
focus on a verification of this concept by the use of older
experimental data taken at JINR synchrophasotron.

An experiment \cite{Baldin} was designed for measurement of
cross-sections of  pp-, ND-, and DD-interactions  at 8.9 GeV
momentum of primary protons and deuterons. A model of the detector
operation was briefly described in \cite{Baldin}. Its parameters
were established by means of measuring differential elastic
cross-sections for proton-proton scattering in a known kinematic
region. Three peaks were observed in the spectrum of the missing
masses of the reaction D$+$D$\to$M$_X+$D at $t=-0.495$ GeV$^2$. Here
we shall concentrate only on one of them, called the third peak in
the original paper. In regard to the third peak, M.A.~Baldin et al
suggested that it might occur due to: a contribution of an excited
state of deuteron; scattering of a constituent quark (entering into
the composition of the incident deuteron) by target deuteron; and,
in addition, N$^*$-baryon production. Experimental findings occurred
after the paper \cite{Baldin} was written give a cause for
re-examination of the suggestions mentioned above.

The present paper might be considered as a particular proposal  for
experimental search of phase transitions in small nucleon systems.

\section{Constituent quark scattering }

Elastic scattering of a constituent quark by the target deuteron may
be considered in the framework of a model in which values of
momentum and mass of the projectile quark are considered in the form
$$
P_q = x P_1, \qquad M_q = x M_D ,
$$
where  $x$ is determined from kinematics of the reaction. A
necessary relation between quark mass and  known experimental
parameters can be found as follows. Let us denote by 1+2 $\to$ 3+4 a
reaction at issue, where the projectile, target and registered
particles are designated by 1, 2 and 4, correspondingly, and 3
denotes an object X which mass should be determined. Two different
expressions for the Lorentz invariant Mandelstam variable $u$,
$u=(p_1 - p_4)^2$ and $u=(p_2 - p_3)^2$, where
$p_i=(E_i,\mathbf{P}_i)$, allow to connect $M_X$ and $\cos \theta$
which  describs the escape direction of the particle 4 in the
laboratory system. A value energy of particle 4 as function of
$M_2,\; M_4$ and $t$ may be found by making use of a relation
$t=(p_2 - p_4)^2$. In addition, $E_3=E_1+E_2-E_4$. Proceeding on
this way, one obtains
$$
M_q  = \frac{{ - M_D^2 t}}{{E_1 t + |\mathbf{P}_1| \sqrt {t( -
4M_D^2 + t)} \cos \theta }},
$$
and $M_q  = 0.351$  GeV for $\cos \theta = 0.396$. This number
contradicts manifestly to estimations of modern  quark models: see,
e.g., \cite{Jovan} where $M_q = 0.318$ GeV. On the other hand, we
shall see below that a peak at $\cos \theta = 0.396$ corresponds
remarkably to the dibaryon found by WASA-at-COSY Collaboration
\cite{WASA}.

\section{ Partial-wave analysis (PWA) and $SU(6)\otimes O(3)$
quark spectroscopy }

Now let us turn to study of a possible contribution  of $N+D \to N^*
+D$ reactions to the experimental cross-section.  Isotopic spin
conservation constrains isospin of $N^*$ to be equal to 1/2.
Therefore, $\Delta-$baryon excitations of nucleon may be ignored
here, and among $N^*$ excitations only N(1440), N(1520) and N(1535)
are important in the kinematic region under consideration. Main
characteristics of the baryon resonances taken into account are
shown in Table~\ref{tab1}.
\begin{table}
\caption{\label{tab1}\small Spin, parity and width of $N^*$ included
in our PWA. The data are given by Particle Data Group \cite{PDG}.}
\begin{center}
\begin{tabular}
{|c |c | c| c| c| l|} \hline
N$^*$ &  $S_{N^*}$ &  $P_{N^*}$ & $\Gamma_{N^*}$, MeV\\
\hline
N(1440) & $1/2$  & $\;$1 & 300  \\
N(1520) & $3/2$  & -1    & 115  \\
N(1535) & $1/2$  & -1    & 150 \\
\hline
\end{tabular}
\end{center}
\end{table}

Besides the spatial parity, $\hat{P}$, conservation, one should
respect the angular momentum,
$\hat{\mathbf{J}}=\hat{\mathbf{L}}+\hat{\mathbf{S}}$, preservation.
In $sp$-approximation, appropriate to the hard collisions, only
$L=0$ and $L=1$ eigenvalues of the orbital momentum can be
considered. In these terms, parities of initial and final states may
be expressed as follows:
\begin{equation}\label{parity}
P_i=P_N P_D (-1)^{L_i} = (-1)^{L_i}= P_f= P_{N^*} (-1)^{L_f} .
\end{equation}

Further PWA may be  simplified essentially via application of the
$SU(6)\otimes O(3)$ description of baryon excitations, suggested by
R.H.~Dalitz and co-authors \cite{Dalitz}. According to it, spin
$\mathbf{S}_{N^*}$ of a nucleon resonance $N^*$ may be represented
as follows:
\begin{equation}\label{spin}
\mathbf{S}_{N^*}=\mathbf{S}_N +\mathbf{l},
\end{equation}
where $\mathbf{S}_N$ is spin of the unexcited nucleon $N$ and
$\mathbf{l}$ is orbital momentum of quarks inside of the excited
nucleon $N^*$. Using (\ref{parity}), it is readily seen that for
each partial wave, which is characterized by fixed values of $J$ and
$P$, a value of parity $P_{N^*}$ of nucleon resonance $N^*$
determines totally a possible behavior of $l$ and $L$ values. For
N(1440), one has $P_{N^*} = P_N = 1$ which implies $l=0$, and,
subject to (\ref{parity}), also $L_f = L_i$. For N(1520) and
N(1535), $P_{N^*} = -1$; therefore $l=1$. According to (\ref{spin})
and Table~\ref{tab1}, we can interpret spins of N(1520) and N(1535)
as two different manners of summation, using Clebsch-Gordan
coefficients,  of quark orbital momentum $l=1$ and initial spin
$S_N$=1/2 of unexcited nucleon. The parity conservation leads to
simultaneous change of $L$ and $l$ values in two possible ways:
\begin{equation}\label{swap}
L_i=1 \to L_f=0, \qquad l_i=0 \to l_f=1,
\end{equation}
and
\begin{equation}\label{init}
L_i=0 \to L_f=1, \qquad l_i=0 \to l_f=1.
\end{equation}
In the frame of $SU(6)\otimes O(3)$ spectroscopy, these cases
correspond to conservation of eigenvalues of operator
$\mathbf{M}^2=(\mathbf{L}+ \mathbf{l})^2$, which are equal to 2 and
0, accordingly. Operator of the total orbital momentum $\mathbf{M}$
commutes with $\mathbf{M}^2$, and we can develop a more detail
picture including account of a direction of $\mathbf{M}$. Below we
shall consider centrally symmetric interaction conserving the
direction of $\mathbf{M}$. In this case, conservation of the total
angular  and orbital momenta implies preservation of the total spin,
$\mathbf{S}=\mathbf{J}-\mathbf{M}$, of the system and our
description admits a further development.

A general expression of the $N+D \to N^* +D$ amplitude linear
relative to $S_N$, $S_D$ and invariant under time reversal and space
rotation or reflection is as follows \cite{Landau}
\begin{equation}\label{ampg}
T(\mathbf{S}_N ,\mathbf{S}_D ) = C_1  + C_2 (\mathbf{S}_N  +
\mathbf{S}_D ) \cdot \mathbf{j}  + C_3 (\mathbf{S}_N  - \mathbf{S}_D
) \cdot \mathbf{j}
\end{equation}
$$
+C_4 (\mathbf{S}_N  \cdot \mathbf{j} )(\mathbf{S}_D \cdot \mathbf{j}
) + C_5 (\mathbf{S}_N \cdot \mathbf{k} )(\mathbf{S}_D  \cdot
\mathbf{k} ) + C_6 (\mathbf{S}_N \cdot \mathbf{i} )(\mathbf{S}_D
\cdot \mathbf{i} ),
$$
where
$$
\mathbf{j}  = \frac{{\mathbf{p} \times \mathbf{p'}}} {{\left|
{\mathbf{p} \times \mathbf{p'}} \right|}}, \qquad \mathbf{k} =
\frac{{\mathbf{p} - \mathbf{p'}}} {{\left| {\mathbf{p} -
\mathbf{p'}} \right|}}, \qquad \mathbf{i}  = \frac{{\mathbf{p} +
\mathbf{p'}}} {{\left| {\mathbf{p} + \mathbf{p'}} \right|}},
$$
$\mathbf{p} $ and $\mathbf{p'}$ are momenta of the ingoing nucleon
and outgoing N$^*$. Here $C_i$ are scalar functions which may depend
only on a scalar $(\mathbf{p} \cdot \mathbf{p'})/ |\mathbf{p} | |
\mathbf{p'} |$ which is in one-to-one correspondence with $\cos
\theta$. In fact, we should claim $C_3=0$, for $\mathbf{S}_N -
\mathbf{S}_D $ does not commute with $(\mathbf{S}_N + \mathbf{S}_D
)^2$ and the corresponding term breaks conservation of an absolute
value of the total spin. Similarly, it is possible to show that
$C_4=C_5=C_6=0$ \footnote{For any vector $\mathbf{n}$ an identity
$(\mathbf{S}_N \cdot \mathbf{n})(\mathbf{S}_D \cdot \mathbf{n}) =
\frac{1}{2}\left( {(\mathbf{S} \cdot \mathbf{n})^2 - (\mathbf{S}_N
\cdot \mathbf{n})^2 - (\mathbf{S}_D \cdot \mathbf{n})^2} \right)$
holds true. The term $(\mathbf{S}_N \cdot \mathbf{n})^2=1/4$ in the
parentheses preserves $\mathbf{S}$, the term $(\mathbf{S} \cdot
\mathbf{n})^2$ commutes with $\mathbf{S}^2 $, but does not with
$\mathbf{S}$. This means that it conserves an absolute value of the
total spin and breaks its direction. The term $(\mathbf{S}_D \cdot
\mathbf{n})^2$ does not maintain a direction of $\mathbf{S}_D$ and
therefore  a direction of $\mathbf{S} =\mathbf{S}_D + \mathbf{S}_N$
or even an absolute value of the total spin. }. Because of the total
spin conservation, a term proportional to $(\mathbf{S}_N +
\mathbf{S}_D )^2$ is not included in (\ref{ampg}) as far as it is
proportional to unit operator for any state with a total spin $S$
fixed (where, as usual, $\mathbf{S} ^2 = S(S+1)$). Efficiently, it
is included in $C_1$.

Thus, we have seen that the parity conservation admits concordant
alteration of $L$ and $l$ according to (\ref{swap}) and
(\ref{init}). From the physical point of view (\ref{swap})
corresponds to swapping external orbital momentum of N$+$D system
into nucleon, and (\ref{init}) corresponds to excitation of the both
external, $ \mathbf{L}$, and intranucleonic, $\mathbf{l}$, momenta.
These processes may be described by a nonlocal operator $(\mathbf{R}
\cdot \mathbf{r})$ included in the interaction amplitude. Here
$\mathbf{R}$ is a polar vector given in the laboratory system, which
is directed at center of inertia of N$+$D system, and $\mathbf{r}$
is a polar vector pointed at center of mass of the  nucleon
colliding with deuteron. Without loss of generality, we may also
suggest $(\mathbf{R} \cdot \mathbf{R}) = (\mathbf{r} \cdot
\mathbf{r}) =1$. Then $T$-matrix describing production of baryon
from Table~\ref{tab1} may be written in the form
$$
T(N + D \to N^*  + D) = A + B(\mathbf{S}_N  + \mathbf{S}_D ) \cdot
\mathbf{j}
$$
\begin{equation}\label{T1}
+(\mathbf{R} \cdot \mathbf{r})\left[ {C + D(\mathbf{S}_N +
\mathbf{S}_D ) \cdot \mathbf{j} } \right].
\end{equation}
Here $A$ and $B$ describe spin independent and spin dependent parts
of interaction corresponding to N(1440) production. Similarly, $C$
and $D$ describe interaction corresponding to N(1520) and N(1535).
Using an identity
$$(\mathbf{R} \cdot \mathbf{r}) = \frac{1} {2}\left( {R_ +  r_ -   + R_ -  r_ +  }
\right) + R_z r_z$$ and well-known formulae for $\mathbf{R}$ and
$\mathbf{r}$ operators \cite{Landau}
$$
\left\langle {L = 1, M = 0} \right|R_z \left| {L = 0, M = 0}
\right\rangle  =  - i/\sqrt 3,
$$
$$
\left\langle {l = 1, m = 0} \right|r_z \left| {l = 0, m = 0}
\right\rangle  =  - i/\sqrt 3,
$$
$$
\left\langle {L = 1, M =  - 1} \right|R_ -  \left| {L = 0, M = 0}
\right\rangle  =  - i\sqrt {2/3},
$$
$$
\left\langle {l = 1, m = + 1} \right|r_ +  \left| {l = 0, m = 0}
\right\rangle  =  + i\sqrt {2/3},
$$
it is possible to find that amplitudes of the processes (\ref{swap})
and (\ref{init}) are equal to $1$ and $1/3$, accordingly.

\section{Observable particles, cross-section}
In fact, baryon resonances N(1440), N(1520) and N(1535) were not
observed directly. They were present in an intermediate state and
may be identified only via their decay products. Therefore
interference terms corresponding simultaneous propagation of matter
through several quantum states with different spins and parities
should be taken into account. We take for granted that possible
final states tolerating macroscopic recognition may contain N$\pi$,
N$\pi \pi$ and N$\eta$, of course, besides deuteron. For N(1440) and
N(1520), corresponding decay probabilities can be estimated as
$w_{1\pi} \approx 0.65$, $w_{2\pi} \approx 0.35$, $w_{\eta} \approx
0$; and $w_{1\pi} \approx 0.5$, $w_{2\pi} \approx 0.1$, $w_{\eta}
\approx 0.4$ for N(1535), see \cite{PDG}.

Baryon resonances leave imprint of their existence  only as
propagators in total amplitude. For example, a transition
N$+$D$\to$N$+\pi+$D is described by the following $T$-matrix:
$$
T(N+D \to N+\pi+D)  =
$$
$$
\frac{{(A + B (\mathbf{S}_N  + \mathbf{S}_D ) \cdot
\mathbf{j})T(N(1440) \to N + \pi ) }} {{M_{N(1440)}^2  - M_X^2 -
iM_{N(1440)} \Gamma _{N(1440)} }}
$$
\begin{equation}\label{ampl}
+  {\frac{{f(S,3/2)\left( {C + D (\mathbf{S}_N  + \mathbf{S}_D )
\cdot \mathbf{j} } \right) T(N(1520) \to N + \pi )}} {{M_{N(1520)}^2
- M_X^2 - iM_{N(1520)} \Gamma _{N(1520)} }}}
\end{equation}
$$
+   {\frac{{f(S,1/2)\left( {C + D (\mathbf{S}_N  + \mathbf{S}_D )
\cdot \mathbf{j} } \right) T(N(1535) \to N + \pi ) }}
{{M_{N(1535)}^2  - M_X^2 - iM_{N(1535)} \Gamma _{N(1535)} }}}.
$$
Analogous expressions take place for N$+$D$\to$N$+\pi+\pi+$D and
N$+$D$\to$N$+\eta+$D transitions. In (\ref{ampl}), scalar functions
$A,\; B,\; C,\; D$ are the same as in (\ref{T1}), and factors
$f(S,N^*)$ may be found on basis of Clebsch-Gordan coefficients, as
it was mentioned in previous section. Following this prescription,
one can find
$$
f(S ,S_{N^* } ) = \sum\limits_{\sigma _1  =  \pm 1/2} {}
\sum\limits_{\sigma _2  = 0, \pm 1} {} \sum\limits_{m = 0, \pm 1}
$$
$$
{\left\langle {{\frac{1} {2}\sigma _1 1\sigma _2 }}
 \mathrel{\left | {\vphantom {{\frac{1}
{2}\sigma _1 1\sigma _2 } {S ,\sigma _1  + \sigma _2 }}}
 \right. \kern-\nulldelimiterspace}
 {{S ,\sigma _1  + \sigma _2 }} \right\rangle } \left\langle {{\frac{1}
{2}\sigma _1 1m}}
 \mathrel{\left | {\vphantom {{\frac{1}
{2}\sigma _1 1m} {S_{N^* } ,\sigma _1  + m}}}
 \right. \kern-\nulldelimiterspace}
 {{S_{N^* } ,\sigma _1  + m}} \right\rangle ,
$$
and
$$
  f\left( {\frac{1}
{2},\frac{1} {2}} \right) = 2 + \sqrt 2 , \;     f\left( {\frac{3}
{2},\frac{1} {2}} \right) = \frac{2} {3},
$$
$$
f\left( {\frac{1} {2},\frac{3} {2}} \right) = 0,   \; f\left(
{\frac{3} {2},\frac{3} {2}} \right) = \frac{4} {3}\left( {\sqrt 2  +
\sqrt 3  + \sqrt 6 } \right) ,
$$
where we adopted notations of the Clebsch-Gordan coefficients from
\cite{Landau}.

Here we should re-arrange a usual formula for cross-section
\cite{Pilk},
$$
\frac{{d^2 \sigma }} {{dt \; dM_X^2 }} = \frac{\pi } {{\lambda
^{1/2} (s,m_N^2 ,m_d^2 )}}\frac{1} {{(2S_N  + 1)(2S_D  + 1)}}
$$
\begin{equation}\label{cross}
\times \sum\limits_{{\rm M}_i,{\rm M}_f} {\int {d{\rm Lips}(M_X^2,\;
decay\; products)} } |T_{{\rm M}_i {\rm M}_f }|^2,
\end{equation}
where ${\rm M}_i$ and ${\rm M}_f$ are spin projections of particles
in initial and final states, into terms of our model of the orbital
nucleon excitations. To this end,  we replace averaging over ${\rm
M}_i$ and summation over ${\rm M}_f$ by corresponding operation over
$\Sigma_i$ and $\Sigma_f$, which are total spin projections of {\it
quarks} in initial and final states. For nonpolarized initial
states, probabilities of occurrence of $S=1/2$ and $S=3/2$ are equal
to $1/3$ and $2/3$, accordingly. Taking into account that the
contribution of orbital excitations is already included by means of
$f(S ,S_{N^* })$, we may write:
$$
\frac{1} {{(2S_N  + 1)(2S_D  + 1)}}\sum\limits_{{\rm M}_i,{\rm M}_f}
|T_{{\rm M}_i {\rm M}_f }|^2
$$
$$
= \frac{1}{3}\sum\limits_{\Sigma_f = \pm 1/2} \overline
{|T_{\Sigma_i \Sigma_f }|^2} + \frac{2}{3}\sum\limits_{\Sigma_f =
\pm 1/2, \pm 3/2} \overline {|T_{\Sigma_i \Sigma_f }|^2},
$$
and then transform
$$
\sum\limits_{\Sigma_f} \overline {|T_{\Sigma_i \Sigma_f }|^2} =
\sum\limits_{\Sigma_f} \overline {T_{\Sigma_i \Sigma_f}
T^*_{\Sigma_i \Sigma_f}} = \sum\limits_{\Sigma_f} \overline
{T_{\Sigma_i \Sigma_f} T^{\dag}_{\Sigma_f \Sigma_i}}
$$
$$
 = \overline{(T T^{\dag})}_{\Sigma_i \Sigma_i} \equiv
\frac{1}{2S+1} {\rm Tr} (T T^{\dag}).
$$
Now it is easy to prove a relation
$$
\frac{1} {{(2S_N  + 1)(2S_D  + 1)}}\sum\limits_{{\rm M}_i,{\rm M}_f}
|T_{{\rm M}_i {\rm M}_f }|^2
$$
$$
= \frac{1}{6}\sum\limits_{S=\frac{1}{2}, \frac{3}{2}} {\rm Tr} (T(S)
T(S)^{\dag}),
$$
which means that values of total spin $S=1/2$ and $3/2$, as well as
all its projections $\Sigma = \pm 1/2$ and $\Sigma = \pm 1/2,\; \pm
3/2$, correspondingly,  give equal contribution to the final result.
It should be stressed that the sign $^{\dag}$ of Hermitian
conjugation refers to $T$  as to spin operator, and it does not mean
transposition of other variables\footnote{This mathematical trick is
described in \cite{Landau} in section devoted to spin-orbit
interaction. }.

Calculations of ${\rm Tr} (T T^{\dag})$ may be completed with making
use of relations: ${\rm Tr}(\mathbf{S}  \cdot \mathbf{j} ) = 0,$
$$
{\rm Tr}(1) = \left\{ {\begin{array}{*{20}c}
   {2, \;\;    S = 1/2,}  \\
   {4, \;\;    S = 3/2,}  \\
 \end{array} } \right. \qquad
{\rm Tr}(\mathbf{S}  \cdot \mathbf{j} )^2  = \left\{
{\begin{array}{*{20}c}
   {1/2,  \;\;  S  = 1/2,}  \\
   {5,    \;\;\;\;\;\;   S  = 3/2.}  \\
 \end{array} } \right. \qquad .
$$

Absolute values of the decay amplitudes are fixed in terms of decay
widths \cite{Pilk},
$$
\Gamma_{N^*,f}=\frac{1}{2M_{N^*}} \int d{\rm Lips}(M_{X}^2,f)
\sum_{{\rm M}_f}|T(N^* \to f)|^2,
$$
where subscript $N^*$ denotes a particular baryon resonance, $f$ is
its decay products. We confine our estimations of interference
between different baryon resonances to operations with phase space
averaged values. For this purpose, we define\footnote{Hereafter we
retain the overline as notation for averaging over Lorentz-invariant
phase space.}
$$
\Gamma_{N^*,f}=\frac{(2S_{N^*}+1)}{2M_{N^*}}\overline{|T(N^* \to
f)|^2} {\rm Lips}(M_{X}^2,f),
$$
and substitute\footnote{Using  Cauchy-Bunyakovsky-Schwarz
inequality, it  may be proven that modulus of the interference terms
defined by (\ref{average}) is in the general case greater than the
true one. Therefore the role of interference is overestimated in our
calculations. Thus, we create an optimum for explanation of
experimental data by interference between different nucleon
excitations, as far as the resonances have too large widths to
explain cross-section by themselves.}
\begin{equation}\label{average}
(2S_{N^*}+1)\Bigl(\overline{|T(N_i^* \to f)|^2}\Bigr)^{1/2}
\Bigl(\overline{|T(N_j^* \to f)|^2}\Bigr)^{1/2}
$$
$$
\times e^{i(\overline{ \alpha}_i - \overline{\alpha}_j )} {\rm
Lips}(M_{X}^2,f) =2 \sqrt {M_i M_j \Gamma_i \Gamma_j }
e^{i(\overline{ \alpha}_i - \overline{\alpha}_j )}
\end{equation}
for
$$
\int d{\rm Lips}(M_X^2, f)\sum_{{\rm M}_f} T(N^*_i \to f)T^*(N^*_j
\to f)
$$
if $M_X$ is greater than  $N^*$ decay threshold and zero otherwise.
Here baryon resonances are different, $i \neq j$, and decay
particles  are the same for the both multipliers under integral
sign.

Strictly speaking, separate control of spin projections of $N^*$ is
not kept in mind in our description, but only projection of total
spin of quarks in the final state of reaction N$+$D$\to$N$^*+$D.
Therefore we should take into account availability of deuteron too
and replace ${\rm M}_f$ with $\Sigma_f$ and $S_{N^*}$ with $S$ in
the previous formulae. Such a treatment may be understood as
summation over quark spin projections inside $N^*$ and spectator
deuteron.  Contribution of orbital excitations into spin projection
of $N^*$ is already included explicitly  by means of $f(S ,S_{N^*
})$, as it was mentioned above. This new interpretation of spin
summation rule is an inevitable corollary of consideration of baryon
as a compound system with its own inner structure.

In the accepted approximation, only phases of the decay amplitudes
$\overline{ \alpha}_i$ may be used as adjustable parameters for
experimental data matching. In addition, eight real numbers
corresponding complex parameters $A,\;B,\;C,\;D$ in (\ref{ampl}) are
brought into play for this purpose. Interference terms corresponding
decays of $N^*$  via $\eta$ are absent since cross-sections of this
channel  are negligible quantities but for one of the resonances
under consideration (see values $w_{\eta}$ in beginning of this
section). The final formula describing the experimental data may be
written in the following form:
$$
\frac{{d^2 \sigma }} {{dt \; dM_X^2 }} =  \frac{\pi } {{6 \lambda
^{1/2} (s,m_N^2 ,m_d^2 )}}
$$
$$
\times \sum\limits_{f,S} \int d{\rm Lips}(M_X^2 ,f) {{{\rm
Tr}(T(S,f)T^{\dag}(S,f)  )}+E },
$$
where $f = N\pi ,N\eta ,N\pi \pi$, $S = \frac{1}{2},\; \frac{3}{2}$
and additional adjustable parameter $E$ describes a contribution of
direct pion production near $M_X^2=1.5 \div2$ GeV$^2$.

\section{Some details of numerical calculations}
To reach an  optimum in describing the experimental data we
minimized total deviation square  for  22 experimental points from
the theoretical curve. Ten central experimental points, as the most
important,  were taken with unit weights and six ones on their left
and six ones on their right were scaled with  0.5 significance.
MAPLE procedure NLPSolve for the local minimum search was used for
optimal selection of theoretical parameters. Several series, each
containing 20 000 different sets of random initial values of
parameters, were generated and only 30 percent of them were finished
without interruption because of very big number of steps towards a
local minimum. Points of the interruptions were considered as local
minima too, because they usually correspond to wanderings along
valleys. Then the best local minimum was taken for each of the
series, and values of objective function corresponding to them were
compared. They turned out to be equal within accuracy of 11 decimal
digits. All the best local optima have demonstrated that
experimental data demand unambiguously:
\begin{equation}
\label{zeros}
|A|=0, \qquad |C|=0.
\end{equation}
This means that phases $\phi_A$ and $\phi_C$ of complex numbers $A$
and $C$ have no impact upon objective function. For removal of
degeneration, we have fixed  $\phi_A=\phi_C=0$ and introduced
condition (\ref{zeros}) explicitly into minimizing functional. Now
the normal mode  of NLPSolve  performance increased up to 55 percent
signalling, nevertheless, that a large degeneration still persisted.
Three series of numerical experiments, containing 100, 1000 and
20$\;$000 events, with random selections of  initial values of the
remaining parameters  were fulfilled. They showed that parameters
$|B|$, $|D|$ and $E$ are identical in all the cases and are
determined with accuracy of 4 and 6 decimal digits already in the
series with 100 and 1000 events. However, all phases  underwent
rather strong changes with growth of statistics, signalling that
minimizing functional remains degenerate with respect to them. Thus,
the optimization problem does not allow us to determine  phases of
parameters $A$, $B$, $C$, $D$,  $T(N_i^* \to N+\pi)$ and $T(N_i^*
\to N+\pi+\pi)$, because many of their sets describe equally well
the experimental data. A grade of fidelity of reproduction of the
experimental data by this model may be seen in Fig.~1.

We  have also fulfilled evaluation of the model parameters using
only 10 experimental points taken straight from   the fine structure
location, trying to enhance an impact of the most important region.
It was technically fully regular procedure, as far as we had only 8
independent parameters at that stage. However, an agreement between
theory and experiment has not been improved  even in this case.

\section{ Conclusions and Discussion }
Numerical analysis fulfilled within the bounds of our model has
revealed two nonobvious properties of hard N-D and D-D scattering.
First of all, it was established that experimental data
\cite{Baldin} show strong spin dependence of N$+$D$\to$N$^*+$D
transition amplitude, see (\ref{T1}) with $A=C=0$. Secondly,
comparison of the experimental data and theory shown in Fig.~1 makes
an explicit hint of dibaryon production in this kinematic region.
\begin{figure}
\begin{center}
\includegraphics[width=8.6cm]{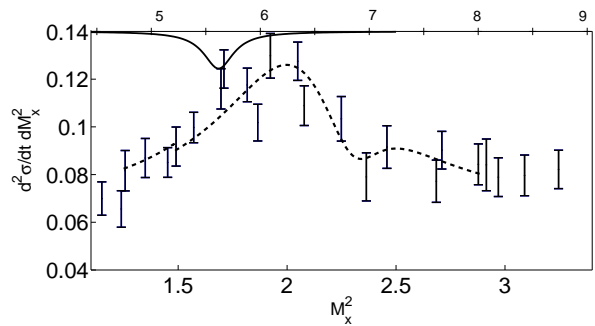}
\end{center}
\caption{\small The experimental data (bars) in the range of the
third peak and their explanation by the sum of contributions of
N$+$D$\to$N$^*+$D reactions (dashed line). The lower scale
corresponds to the kinematics of reaction N$+$D$\to$N$^*+$D, the top
one describes reactions D$+$D$\to$X$+$D, which implies the dibaryon
production.  A possible contribution of a dibaryon at 2.37 GeV,
$\Gamma\approx 70$ MeV reported by WASA-at-COSY Collaboration
\cite{WASA} into the cross-section is shown with the overturned
solid line.}\label{Peak3}
\end{figure}
Indeed, on the one hand, consideration only usual nucleon
excitations cannot explain the fine structure shown in the figure.
On the other hand, assumption about presence of a dibaryon at
$M_{2B} \approx  $ 2.38 GeV, $\Gamma_{2B}\approx$ 70 MeV,  seen by
WASA-at-COSY Collaboration \cite{WASA} allows one to explain it very
naturally.  Isospin conservation  predicts certainly that reaction
D$+$D $\to$ dibaryon $+$ D should yield dibaryon with isospin $I=0$,
which also corresponds to the WASA-at-COSY result \cite{WASA}. Thus,
our consideration of the data on the hard deuteron-deuteron
scattering \cite{Baldin}   meets the expectation to observe the
transition of nucleon matter into other states using the method of
cumulative particle which allows to recognize quasi-resonance peaks
in the reaction cross-section.

To check our conclusions, it would be enough to measure with a good
precision production cross-sections of N(1440), N(1520) and N(1535)
from N$+$D $\to$ N$^* +$D reactions  in appropriate kinematic
region, and direct production of pions therein. This allows one to
take into account the background. In addition, repeating experiment
\cite{Baldin} with higher accuracy is necessary too for unambiguous
recognition of dibaryon  by its mass and width. Theoretical and
experimental study of the phases entering into expression for
production amplitude is ineffectual in this respect, so long as
resultant cross-section is weakly dependent on them (see previous
section). Investigation of  decay products of dibaryon will make it
possible to identify its spin and parity and compare with $J^P =
3^+$ observed in \cite{WASA}.

It is interesting to review  ability of the lattice QCD to say
something definite about existence of dibaryons. All lattice QCD
collaborations  have found stable NN-dibaryons and dibaryons
containing s-quarks, but quark masses in their calculations are
higher than  the physical values, see, e.g., \cite{HAL, NPLQCD}.
Chiral extrapolations of these results to the physical point gave,
however, evidences against the existence of such dibaryons, see,
e.g., \cite{Shana}. These calculations deal with ground states and
say nothing about unstable states corresponding to a possibility of
two-baryon fusion into 6-quark bag with a value of mass larger than
a sum of masses of the initial baryons. Recent progress in excited
baryon spectroscopy is depicted in \cite{Lin, Edw}. Corresponding
results based on nonphysical quark masses too cover only one-baryon
states so far and are in a poor agreement with experimental N and
$\Delta$ excitation spectra. The first excited state in two-nucleon
system was found in lattice QCD in \cite{PACS} but with a heavy
quark mass corresponding to $m_{\pi}=0.8$ GeV. Therefore, predicting
quasi-bound states of a multibaryon systems remains a difficult
challenge in lattice QCD till now.

Another important question: what is the reason that so few signs of
dibaryons currently exist in spite of  their search  in the network
of partial-wave analysis? The most likely answer, as we see it, is
still an unsatisfactory precision of PWA. Indeed, incorporating the
additional data of WASA-at-COSY Collaboration into the SAID analysis
produces a pole in support of the resonance hypothesis \cite{SAID}.

A trivial generalization of the method of a cumulative particle is
to select events with several, $n>1$, secondary particles, not
necessarily containing a cumulative one, which accompany a dibaryon
production. Such a group of the additional particles, e.g., pions,
may take away an excess of excitation energy, which put the main
obstacle in the way of dibaryon recognition. In particular, Yu.A.
Troyan reported the registration of some dibaryons using just this
method \cite{Troyan1, Troyan2, Troyan3} It should be, however, noted
that most of experimental searches of dibaryons carried out in the
past must be  exposed to requalification. Let us consider, for
example, a paper by B.M.Abramov et al \cite{Abramov}, which  is
cited sometimes  as a convincing argument against one of Yu.A.
Troyan's experiments. Even gross inspection of that paper reveals
the following grave shortcomings. Firstly, no methods  of a
background substraction have been used. The solid line in the main
figure of the paper \cite{Abramov} is only  an optimal approximation
of the experimental invariant mass spectrum containing, in the
general case, a sum of background and dibaryon contributions.
Secondly,  number of events and precision of measurements do not
allow to obtain a mass spectrum resolution nearly 1 MeV, which is
necessary to verify confidently the Troyan results. Thirdly, the
conditions of "deep cooling", which was ensured in Troyan's
experiments, has not been fulfilled in \cite{Abramov} (see
\cite{Kostenko3} for details).

All dibaryons reported in \cite{Troyan1, Troyan2} were observed in
inelastic N-N interactions with additional secondary pions.  The
extra pions  take away an excess of excitation energy  -- a process
which is a some kind of annealing. This may reconcile two opposite
requirements imposed simultaneously on the system: it must be
strongly compressed to form a compound state and it must be cold
enough, since highly excited levels are usually short-living and
elusive. Two additional pions in final state were  in WASA-at-COSY
and CELSIUS/WASA Collaborations experiments \cite{WASA, WASA1,
CELSIUS}. Therefore, we may suggest with high reliability that
synthesis of new multibaryons, and particularly dibaryons, should
succeed an observation made also for synthesis of new transuranium
elements -- the system must be as much cold as possible to be
observable readily.

\begin{acknowledgments}  We are obliged to R.L.~Workman for his letter
illuminating the current interrelation between PWA and the
WASA-at-COSY dibaryon. We are grateful to N.B.~Bogdanova and
A.P.~Ierusalimov for a useful software support \cite{Bogdan, Ierus}
of our investigation.

\end{acknowledgments}

\end{document}